\documentclass[pre,twocolumn,showpacs]{revtex4}
\usepackage{amssymb}
\usepackage{graphicx}
\usepackage{amsmath}

\newcommand{\beq}{\begin{equation}}
\newcommand{\eeq}{\end{equation}}
\newcommand{\bea}{\begin{eqnarray}}
\newcommand{\eea}{\end{eqnarray}}
\begin{document}
\input{epsf}
%
\title{The easy axis of the magnetic response of the conduction electrons
of yttrium and rare earths}
\author{V.Thakor\dag, J.B.Staunton\dag, J.Poulter\ddag, S.Ostanin\dag,
B.Ginatempo\P, and Ezio Bruno\P}
\affiliation{\dag\ Department of Physics, University of Warwick, 
Coventry CV4~7AL, U.K.}
\affiliation{\ddag\ Department of Mathematics, Faculty of Science,
Mahidol University, Bangkok 10400, Thailand.}
\affiliation{\P\ Dipartimento di Fisica and Unita INFM, Universita di Messina, Italy}
\date{\today}
\begin{abstract}
We describe a scheme for first-principles calculations of the static, 
paramagnetic, spin susceptibility of metals with relativistic 
effects such as spin-orbit coupling included. This gives the direction for
an applied modulated magnetic field to maximise its response.
This {\it easy axis} depends on the modulation's wave-vector ${\bf q}$. For 
h.c.p.  yttrium we find the peak response at a ${\bf q}= (0,0,0.57)\pi/c$,
coincident with a Fermi surface nesting vector, to have an easy axis perpendicular to 
${\bf q}$. This is consistent with the {\it helical} anti-ferromagnetic order 
found in many dilute rare earth-$Y$ alloys. Conversely, the easy axis for the 
response to a uniform magnetic field lies along the c-axis. The conduction electrons' 
role in the canting of magnetic moments in $Gd-Y$ alloys and other rare earth 
materials is mooted.
\end{abstract}
\maketitle
Conduction electrons play an important role in the magnetism of rare earths. 
Specifically they determine the Ruderman-Kittel-Kasuya-Yosida (RKKY) indirect 
exchange interactions which connect the localised 4f magnetic moments of the 
materials.~\cite{RKKY} The magnetic response of the conduction electrons, 
the wave-vector 
dependent susceptibility $\chi({\bf q})$, governs the ordering of the moments. 
If the greatest value of $\chi({\bf q})$ is for ${\bf q}=(0,0,0)$, the moments 
order ferromagnetically. A peak for $\chi({\bf q})$ at some finite ${\bf q}$,
on the other hand, signifies a more complicated magnetic structure with a 
modulation wave-vector equal to ${\bf q}$. In many of the heavier rare earths, 
e.g. Tb, Dy, Ho and Er the incommensurate anti-ferromagnetic (AF) ordered 
states, modulated along the c-axes of these h.c.p. crystalline elements, 
depend strongly on the Fermi surface topology of the conduction electrons. 
The salient part of the Fermi surface (FS) that causes this anisotropy is the 
{\it webbing} feature that contains flat parallel sheets perpendicular to the 
$c$-axis forming a strong nesting effect~\cite{Keeton_Loucks,Loucks}.

The link between FS nesting and the ${\bf q}$-dependence of $\chi({\bf q})$ is 
well-known and has been explored by electronic structure 
calculations.~\cite{Liu+Yttium_chi}
There is however another anisotropy in the interaction between the rare earth (RE) 
magnetic moments that we suggest is also traceable to $\chi({\bf q})$. This comes from 
spin-orbit coupling effects on the conduction electrons and affects whether the 
localised moments lie perpendicular to the c-axis and thus form helical AF 
magnetic states or instead have components along the c-axis and order into b-axis 
modulated AF phases. Of course the anisotropic crystal field acting on the 
4f-shell plays an important role in this aspect but the similarity between Gd 
alloys in which the crystal field effects are small and alloys of Dy,Tb, etc 
suggests a generic feature from the conduction electrons. This is the issue we 
address in this letter.  We describe a `first-principles' theoretical formalism 
for the magnetic response of paramagnetic metals in which all relativistic 
effects such as spin-orbit coupling are included. For the first time the {\it 
easy axis} and its dependence upon wave-vector ${\bf q}$ for the magnetic response 
can be calculated. In effect we elucidate a relativistic RKKY 
interaction.~\cite{jbs-rkky}

We present specific calculations of the magnetic response of the transition 
metal yttrium. With the same h.c.p. crystal structure and electronic configuration
as the rare earths, apart from the f-electrons, it provides an excellent model
for their conduction electrons. Much is being learnt about the magnetism of rare earths
from the varied and complex magnetic structures that they form in multilayers with 
yttrium.~\cite{Majkrzak} These range from canted, antiphase domain structure in $Gd/Y$
multilayers to coherent, incommensurate magnetic helices found in $Dy/Y$ and $Ho/Y$
structures. The mediating role of the conduction electrons of $Y$ for these rare earth
magnetic interactions is crucial in these tailored systems.~\cite{Yafet}
High quality samples of yttrium are more easily
come by than the heavy rare earths and recent experiments~\cite{Dugdale} have 
measured directly the `webbing' feature of its Fermi surface that it has in 
common with several rare earths. Moreover the nature of the magnetic ordering of low 
concentrations of rare earth dopants in $Y$ can be directly related to this 
study.~\cite{YGd,Child,Bates}

We begin by considering a paramagnetic metal subjected to a small, external, 
inhomogeneous magnetic field, $\delta {\bf b}^{ext}({\bf r})$, and obtain 
an expression for the induced 
magnetisation $\delta {\bf m} ({\bf r})$. We use relativistic density functional 
theory (RDFT)~\cite{Rajg+Call} to treat the interacting electrons of
the system and derive an expression via a variational linear response 
approach~\cite{Vosko+Perdew,Applied_to_TM,staun1}. Although there 
are a number of non-relativistic studies 
of this type~\cite{SW+Sav} here we include relativistic effects and pay 
particular attention to the magnetic anisotropy of the response. 
From our RDFT starting point we make a  Gordon decomposition of the current 
density~\cite{Rajg+Call} and retain the spin-only part of the current, namely the spin
magnetisation ${\bf m}({\bf r})$. 
This results in a $\emph{spin-only}$ version of 
RDFT.~\cite{Rajg+Call} in which the self-consistent solution of 
Kohn-Sham-Dirac equations is sought, i.e.  
\begin{eqnarray} 
\lbrack\,c \mbox{\boldmath${\tilde{\alpha}}$}.{\bf \hat{p}} + \tilde{\beta}mc^{2} +
\,{\bf \tilde{1}}V^{eff}[\rho,{\bf m}] 
- \tilde{\beta} 
\mbox{\boldmath${\tilde{\sigma}}$}. {\bf b}^{eff}[\rho,{\bf m}] 
- \varepsilon \,\,\rbrack
\nonumber \\
\times \,\,\,  
G({\bf r} , {\bf r}' ; \varepsilon) = {\bf \tilde{1}} \delta ({\bf r}-{\bf r}') 
\,\,
\label{eq:RelKSG}
\end{eqnarray} 
which describes the motion of a single electron through effective fields and
$\mbox{\boldmath${\tilde{\alpha}}$}$ and $\tilde{\beta}$ are Dirac $4 
\times 4$ matrices.
$G({\bf r} , {\bf r}' ; \varepsilon)$ is the one electron Greens function and
the charge $\rho ({\bf r})$ and magnetisation densities 
${\bf m}({\bf r})$ can be written in terms of it i.e.
\begin{equation}
\rho ({\bf r}) = -Tr. \int d \varepsilon f(\varepsilon, \mu,T) \frac{Im}{\pi} 
G({\bf r} , {\bf r}; \varepsilon) \nonumber
\end{equation}
\begin{equation}
{\bf m}({\bf r}) =- Tr. \tilde{\beta} \mbox{\boldmath${\tilde{\sigma}}$} \int d \varepsilon f(\varepsilon, \mu,T) \frac{Im}{\pi}
G({\bf r} , {\bf r}; \varepsilon)
\end{equation}
where $\mu$ is the chemical potential, $T$ the temperature and 
$f(\varepsilon,\mu,T)$ the Fermi-Dirac function. These expressions can be converted 
into sums over fermionic Matsubara frequencies $\omega_n =  i (2n+1) \pi k_B
T$ ~\cite{FW}. 
The effective potential $V^{eff}[\rho,{\bf m}]$ consists of the usual combination
of external potential (from the lattice of nuclei), the Hartree potential and 
functional derivative of exchange-correlation energy $E_{xc} [\rho,{\bf m}]$
with respect to $\rho$ whilst the effective magnetic field 
${\bf b}^{eff}[\rho,{\bf m}]$ is the sum of any external magnetic field (from
magnetic impurities, for example) and the functional derivative of
 $E_{xc}$ with respect to magnetisation. We use the local density approximation (LDA)~\cite{vBH} for $E_{xc}$.  
The leading relativistic effects contained in the Kohn-Sham-Dirac hamiltonian
 of eq.(\ref{eq:RelKSG}) are the well-known
mass-velocity, Darwin and spin-orbit coupling effects. 
  
If a small external field $\delta {\bf b}^{ext}$ is applied along a direction $\hat{{\bf n}}$ 
with respect to the crystal axes of a paramagnetic system, a small magnetisation, $\delta {\bf m} ({\bf r})$, and 
effective magnetic field, $\delta {\bf b}^{eff}$ are set up. The effective magnetic field is 
given by $\delta {\bf b}^{eff}[\rho({\bf r}),{\bf m}({\bf r})] = 
\delta{\bf b}^{ext}({\bf r}) + I_{xc}({\bf r})\delta {\bf m}({\bf r})$ 
where  $I_{xc}({\bf r})$ is the functional derivative of the effective exchange
 and correlation 
magnetic field (within the LDA) with respect to the induced magnetisation density.
The Green's function satisfying equation (\ref{eq:RelKSG}) can be expanded in a 
Dyson equation in terms of the unperturbed Green's function, 
$G_o({\bf r} , {\bf r}' ; \varepsilon)$ 
of the paramagnetic system ($\delta{\bf b}^{eff}$ = 0) and perturbation 
$\tilde{\beta} \mbox{\boldmath${\tilde{\sigma}}$}. \delta{\bf b}^{eff}$ and the 
first order terms enable the magnetic response function to be obtained. 
  
For a general crystal lattice with $N_s$ atoms located at positions ${\bf a}_l$ 
($l$ = 1,..,$N_s$) in each unit cell, a lattice Fourier transform can be carried out 
over lattice vectors $\{{\bf R}_i\}$. This can be written
\begin{eqnarray} 
\chi^{{\hat{\bf n}}} ({\bf x}_{l},{\bf x}'_{l'}\, ,{\bf q}) \, = \,
\chi_{o}^{{\hat{\bf n}}} ({\bf x}_{l},{\bf x}'_{l'}\, ,{\bf q})  
\qquad \qquad  \nonumber \\
+ \,\, \sum_{l''}^{N_{s}} \int \, 
\chi_{o}^{{\hat{\bf n}}} ({\bf x}_l,{\bf x}''_{l''}\, ,{\bf q})
\, I_{xc}({\bf x}''_{l''}) \,
\qquad  \nonumber \\
\times \,\,\,\, \chi^{{\hat{\bf n}}} ({\bf x}''_{l''},{\bf x}'_{l'}\, ,{\bf q})
\, d{\bf x}''_{l''}  
\label{eq:X}
\end{eqnarray} 
where the ${\bf x}_l$ are measured relative to the positions of atoms centred on ${\bf a}_l$. 
The non-interacting susceptibility of the static unperturbed system is given by
\begin{eqnarray} 
\chi_o^{{\hat{\bf n}}} ({\bf x}_l,{\bf x}'_{l'}\, ,{\bf q}) \, = 
\qquad \qquad \qquad \qquad \qquad \qquad \,\,\,\,\,\,\,\,\,\,  
\nonumber \\  
\,\,\,\,
-(k_B T) \,Tr \, \tilde{\beta}\, \mbox{\boldmath${\tilde{\sigma}}$}\cdot{\hat {\bf n}} \, 
\sum_n \int \, \frac{d{\bf k}}{\nu_{BZ}}  
G_o({\bf x}_l,{\bf x}'_{l'}\,,{\bf k}, \mu+i\omega_n)
\nonumber \\ 
\,\,\,\,\,\, 
\times \,\,\, 
\tilde{\beta}\, \mbox{\boldmath${\tilde{\sigma}}$}\cdot{\hat {\bf n}} \,\,
G_o({\bf x}'_{l'},{\bf x}_{l}\,,{\bf k}+{\bf q}, \mu+i\omega_{n})  
\label{eq:Xo}
\end{eqnarray} 
The integral is over the Brillouin zone with wave vectors ${\bf k}$, ${\bf q}$ 
and ${\bf k}+{\bf q}$ within the Brillouin zone of volume $\nu_{BZ}$. The sum is 
over the fermionic Matsubara frequencies. 
The Green's function for the unperturbed, paramagnetic system containing the band structure effects is obtained via relativistic multiple scattering 
(Korringa-Kohn-Rostoker, KKR) theory~\cite{Faulkner+Stocks}. 
We solve
equation (\ref{eq:X}) using a direct method of matrix inversion. The full Fourier
transform is then generated 
\begin{eqnarray} 
\chi^{{\hat{\bf n}}} ({\bf q}) \, = \,
(1/V)\sum_{l} \, \sum_{l'}\, e^{i{\bf q}.({\bf a}_{l}-{\bf a}_{l'}')} \,
\qquad \qquad  \nonumber \\
\times \,\,\,
\int d {\bf x}_{l} 
\int d {\bf x}'_{l'} \,
e^{i{\bf q}.({\bf x}_{l}-{\bf x}_{l'}')} \,
\chi^{{\hat{\bf n}}} ({\bf x}_{l},{\bf x}'_{l'}\, ,{\bf q},\theta,\varphi) 
\label{eq:Final_X}
\end{eqnarray} 
where $V$ is the volume of the unit cell. Some aspects of the numerical methods used to 
evaluate equations (\ref{eq:X}-\ref{eq:Final_X}) of this type can be found 
in~\cite{staun1}. Note that this expression for the 
non-interacting susceptibility 
can be shown to be formally equivalent to one of the familiar type
\begin{eqnarray}
\chi_o^{{\hat{\bf n}}}({\bf q}) \, \propto
\int d{\bf k} \sum_{j,j^{\prime}} \frac{|M({\bf k},
{\bf k}+{\bf q}, \hat{\bf n})|^2 f_{{\bf k},j}(1-f_{{\bf k}+{\bf q},j^{\prime}})}
{\varepsilon_{j^{\prime}}({\bf k}+{\bf q}) - \varepsilon_{j}({\bf k})}
\end{eqnarray}
where $j$ is an electronic band index, $\varepsilon_{j}({\bf k})$ a single
electron energy, $M$ is a matrix element and 
$f_{{\bf k},j}= f(\varepsilon_{j}({\bf k}),\mu,T)$,  the Fermi Dirac function.

The important feature of the response function (equations 
\ref{eq:X}-\ref{eq:Final_X}), is its dependence on the direction of 
the magnetic field, ${\hat{\bf n}}$, which vanishes when relativistic, 
spin-orbit coupling effects are omitted. The direction, 
${\hat{\bf n}}$=(sin$\,\theta$cos$\,\varphi$,sin$\,\theta$sin$\,\varphi$,cos$\,\theta$) 
is defined by polar and azimuth angles ($\theta,\varphi$). 
In an h.c.p system such as $Y$ or $Sc$, if an external magnetic field is 
applied along the ${\hat{\bf n}}=(0,0,1)$ direction, i.e. the c-axis, 
where ($\theta,\varphi$)~$\to$~(0,0), $\chi_o^{z}$ is produced. 
On the other hand $\chi_o^{x}$ is the 
response of the system when the field is applied 
in the $ab$-plane, ${\hat{\bf n}}=(1,0,0)$, and 
($\theta,\varphi$)~$\to$~($\pi/2$,0). We obtain an anisotropy as the difference 
in the  non-interacting susceptibility when an external magnetic field is 
applied in two directions with respect to the crystal axes, 
i.e. ($\chi_o^{{\hat{\bf n}}} - \chi_o^{z}$). This is enhanced by exchange and
correlation effects (eq.\ref{eq:X}). The approach 
presented here is applicable to ordered compounds and elemental metals and can 
be modified to study disordered alloys~\cite{staun1} owing to its 
KKR multiple-scattering framework. In order to gauge the importance of these 
relativistic effects with atomic number, we compare our calculations of the 
magnetic response of $Y$ with its lighter 3d counterpart $Sc$.

We use atomic sphere approximation (ASA), effective one-electron potentials
and charge densities in the calculations for $Y$ and $Sc$ with
experimental lattice constants $a =6.89$, $c=10.83$ 
and $a=6.24$, $c=9.91$ respectively, in atomic units, $a_{0}$.~\cite{Pearson} 
The details of the 
electronic structures using a fully relativistic KKR method compare well with 
those from full potential calculations.~\cite{Wien97} The Fermi surface for $Y$ 
in the H-L-M-K plane is shown in Fig. 1(a). It shows two relatively flat parallel
sheets. The nesting vector ${\bf q}_{inc} = \pi/c(0,0,0.57)$ is indicated by the 
arrow.
Fig. 1(a) is in very good agreement with previous calculations~\cite{Loucks}
and experiments~\cite{Dugdale,Vinokurova}.
Dugdale {\it et al}.~\cite{Dugdale} have recently carried out positron 
annihilation fermiology
experiments and measured ${\bf q}_{inc} = \pi/c(0,0,0.55 \pm 0.02)$
for $Y$. Also, Vinokurova {\it et al}.~\cite{Vinokurova}, measured
${\bf q}_{inc} \approx \pi/c(0,0,0.58)$. We find a similar Fermi surface for Sc 
with the same
nesting vector ${\bf q}_{inc} = \pi/c(0,0,0.57)$, also in good agreement with earlier
calculations~\cite{FlemLoucks}.

\begin{figure}
\begin{center}
\qquad  \,\,\,
\scalebox{0.45}[0.45]{\includegraphics{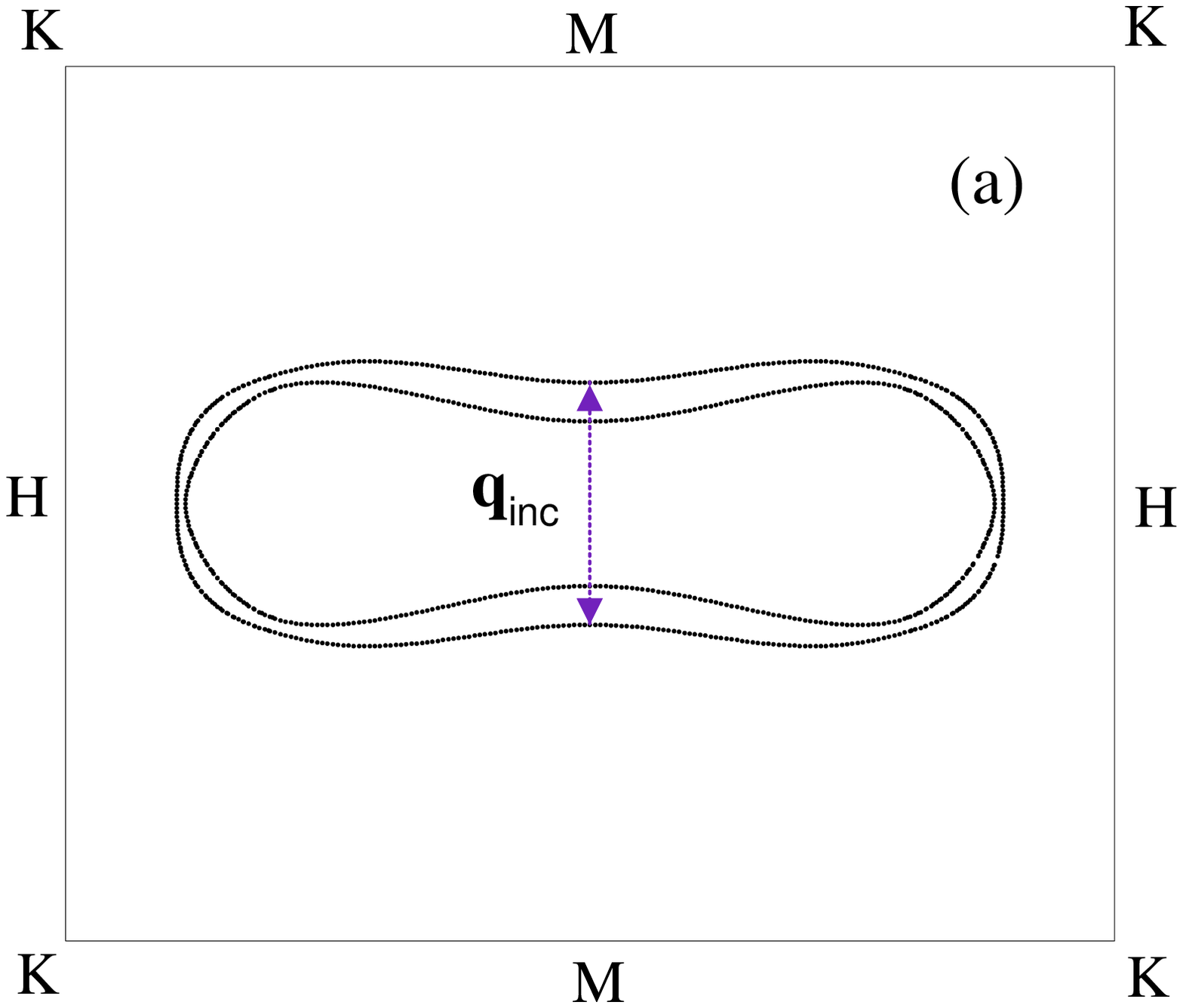}}
\end{center}
\begin{center}
\scalebox{0.48}[0.48]{\includegraphics{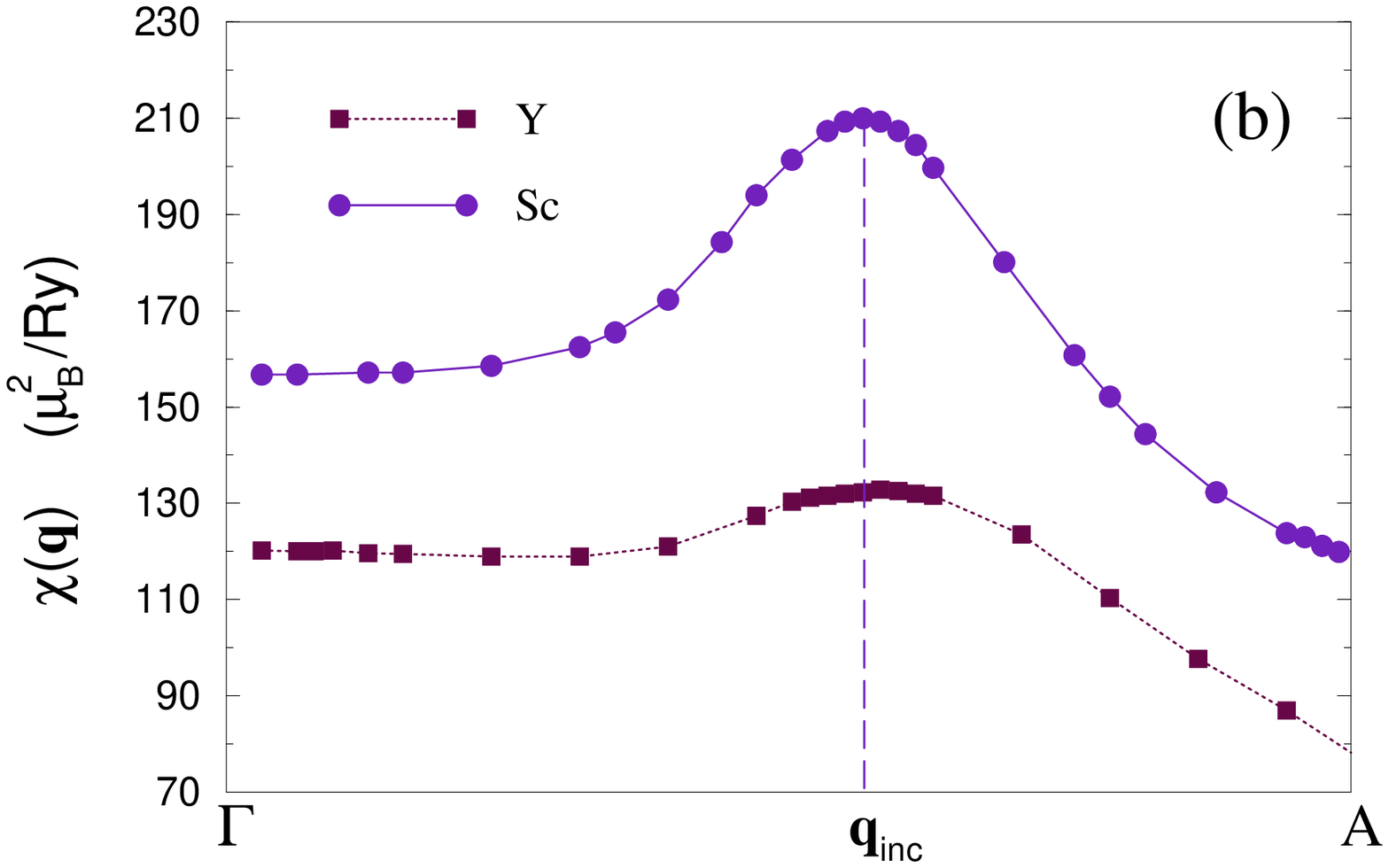}}
\end{center}
\caption{(a) Cross-section of the `webbing' Fermi surface in the H-L-M-K plane for 
$Y$. The nesting vector is indicated by the arrow. The special point L lies 
at the centre.(b) 
The enhanced susceptibility for $Y$ and $Sc$ along the $\Gamma$-A direction for
 T = 100K.} 
\label{FermiSurface}
\end{figure}

Our calculations of the enhanced static susceptibility, defined by equation
 (\ref{eq:X}), 
show a peak at this same wave-vector ${\bf q}_{inc} = \pi/c(0,0,0.57)$ for both $Y$
 and $Sc$. This
is shown in Fig. 1(b), where we probe wave-vectors along the $c$-axis from $\Gamma$ to A. 
(The special point A, for h.c.p. crystal structures is $(0,0,\pi/c)$.) 
The temperature is set at 100K for these calculations.

\begin{figure}
\begin{center}
\qquad  \,\,\,
\scalebox{0.44}[0.44]{\includegraphics{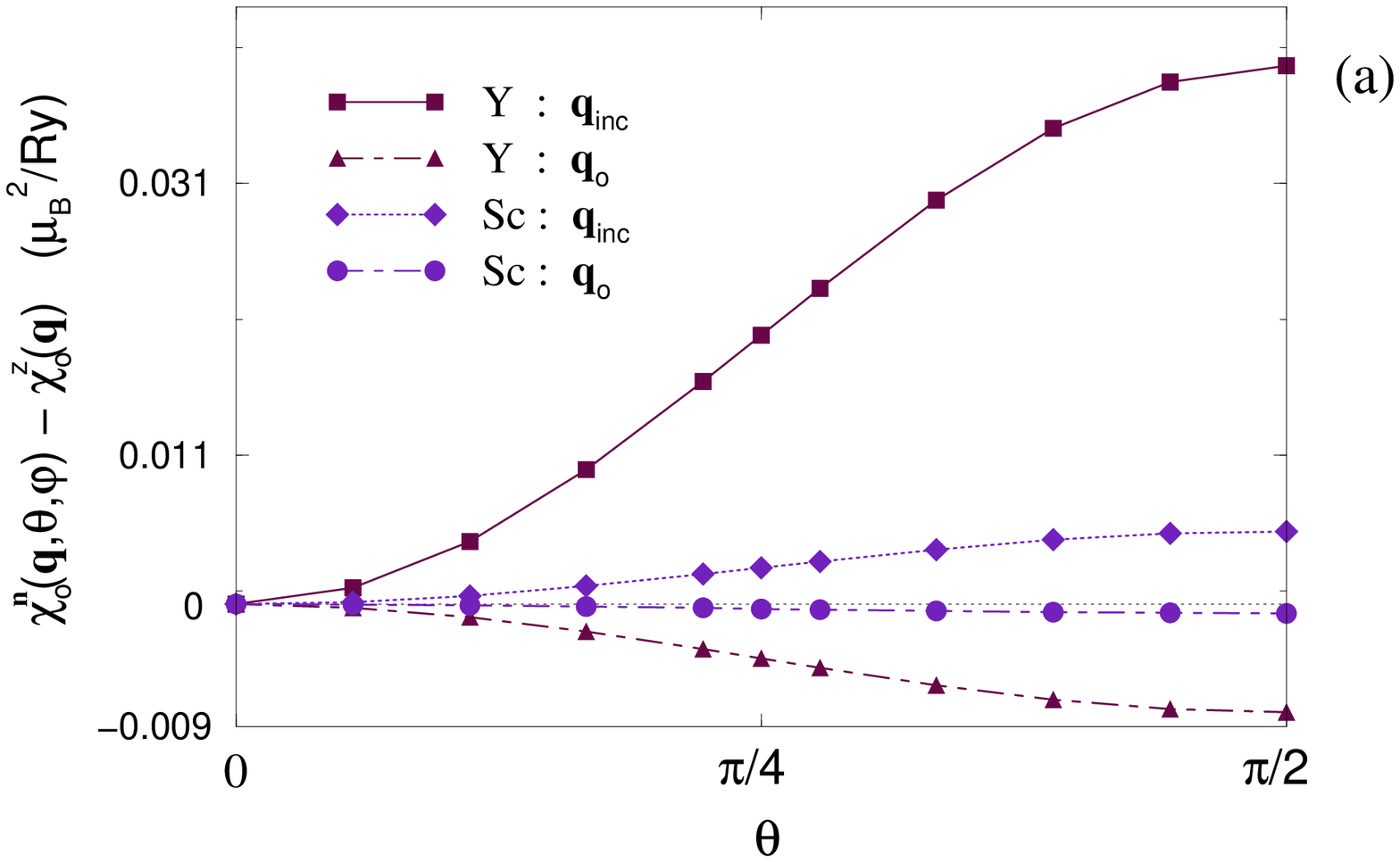}}
\end{center}
\begin{center}
\scalebox{0.44}[0.44]{\includegraphics{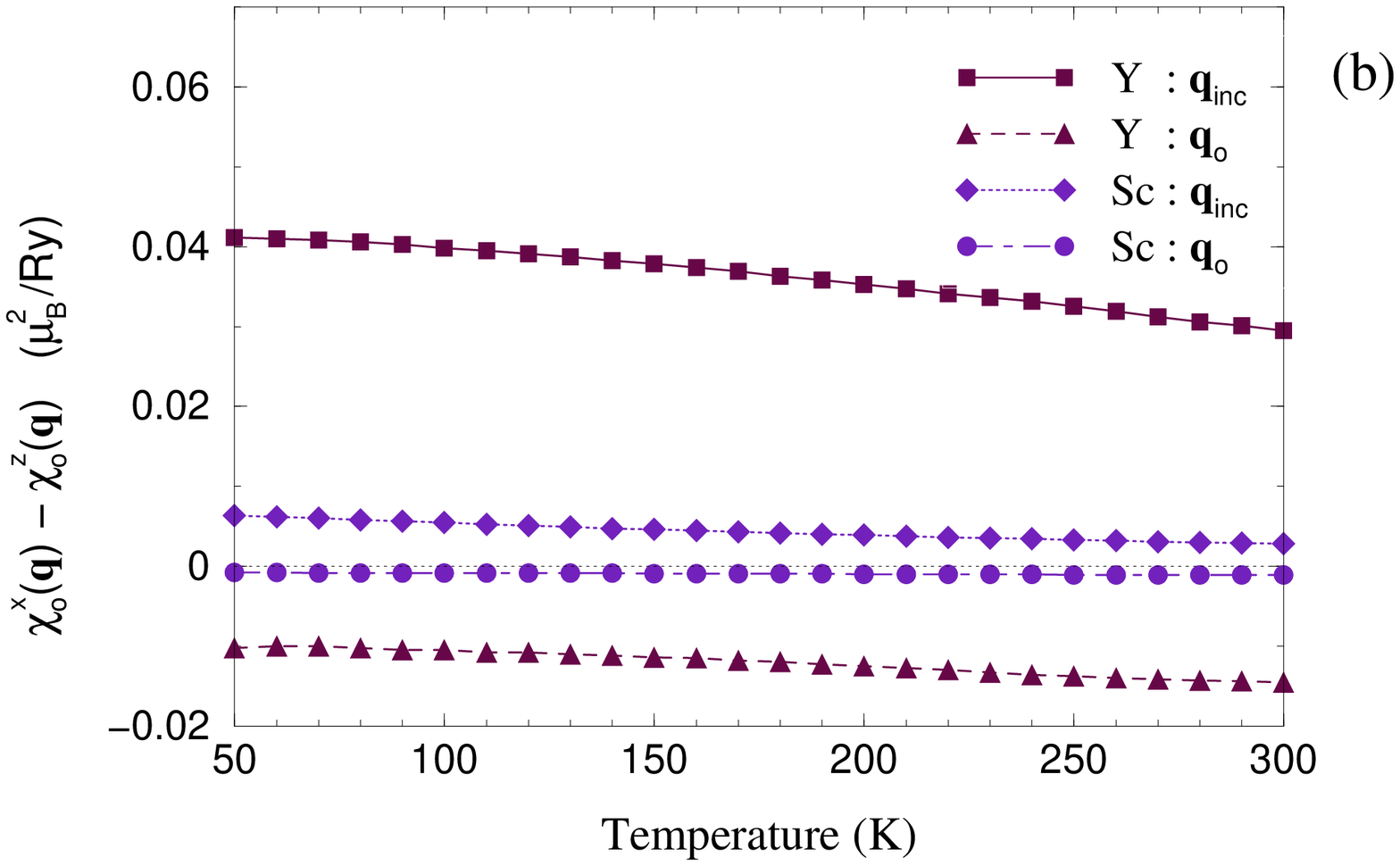}}
\end{center}
\caption{(a) The anisotropy as a function of $\theta$ for $Y$ and $Sc$ at  
${\bf q}_{inc}=\pi/c(0,0,0.57)$ and ${\bf q}_{o}\simeq (0,0,0)$ and at T = 100K.  
(b) The temperature dependence of the anisotropy of $Y$ and $Sc$ at both 
the nesting vector ${\bf q}_{inc}$ and ${\bf q}_{o}$.}
\label{Anisotropy_Y}
\end{figure}

Fig. 2(a) shows the anisotropy $\chi_o^{{\hat{\bf n}}}({\bf q},\theta,\varphi)-\chi_o^{z}({\bf q})$ as a function of $\theta$ for $Y$ and $Sc$ at ${\bf q}_{inc}$ and
${\bf q}_{o} \simeq (0,0,0)$. As expected on symmetry grounds, we find the 
anisotropy to be 
invariant in $\varphi$. This means that the magnetic response is insensitive
to the direction along which an external magnetic field is applied in the ab-plane.
 As $\theta$ is increased, an anisotropy is observed
which reaches a maximum at $\theta = \pi/2$ as shown in Fig. 2(a). In fact, we observe 
that the anisotropy takes the rather simple form
\begin{equation} 
\chi_o^{{\hat{\bf n}}}({\bf q},\theta,\varphi)-\chi_o^{z}({\bf q}) 
= ( \chi_o^{x}({\bf q}) - \chi_o^{z}({\bf q})) \cdot
 |{\hat{\bf q}} \times {\hat{\bf n}}|^2
\end{equation}
 for h.c.p. crystal structures~\cite{Remark} ($|{\hat{\bf q}} \times {\hat{\bf n}}|=
\sin \theta$). It is therefore sufficient to show 
$\chi_o^{x}({\bf q})-\chi_o^{z}({\bf q})$, to determine the easy axes of the 
magnetic response. 
Fig. 2(b) shows the weak temperature dependence of the anisotropy of $Y$ and $Sc$ 
at both the nesting vector 
${\bf q}_{inc}$ and also at ${\bf q}_o \simeq (0,0,0)$.  As expected, the
anisotropy is an order of magnitude larger for $Y$ than for $Sc$. This 
difference is a result of spin-orbit
coupling being more pronounced in the heavier 4d metal $Y$ than in the 3d $Sc$. 
We infer that a still greater but similar anisotropy  should be evident in the 
magnetic response of the conduction electrons in the heavier still RE materials. 

It is apparent from Figs.1 and 2 that $Y$ shows its strongest response
at the wave-vector ${\bf q}_{inc}$ which corresponds to the FS
nesting and that the response here is strongest to magnetic fields directed in 
the basal $ab$-plane. Here, then, is an explanation as why $Gd$ impurities, 
 denoted typically as S ions (zero angular momentum) and thus free from crystal field effects, order into
{\it helical} incommensurate AF states.~\cite{YGd} It is also a strong factor for the 
similar magnetic order found in other dilute rare-earth alloys such as 
$\underline{Y}Tb$, $\underline{Y}Dy$ and $\underline{Y}Ho$.~\cite{Child} 

Fig.1(b) shows that yttrium's responses to ferromagnetic ($\chi({\bf q}) 
= (0,0,0))$) and ${\bf q}_{inc}$-AF modulations are close in strength. 
Figs.2(a) and 2(b) demonstrate, however, that the easy axis for a FM modulation 
lies along the c-axis and therefore contrary to that of the AF one. These features
are pertinent to the magnetic structure of $Gd-Y$ alloys. Pure gadolinium has a  
ferromagnetic phase with the easy axis along the c-axis so that in the paramagnetic
phase at higher temperatures $\chi({\bf q})$ peaks at ${\bf q}= (0,0,0)$. Moreover
its FS here  has been observed not to possess a 
webbing feature \cite{Fretwell}. Some recent experiments~\cite{Fretwell} have shown 
that adding more than around 30$\%$ $Y$ to $Gd$ changes the topology of the 
Fermi surface of the
paramagnetic phase and the webbing is observed again. This is coincident with the
alloys' forming helical AF states at lower temperatures. Moreover, in this 
concentration range, neutron-diffraction~\cite{Bates,Fretwell} has shown that 
application of a modest uniform magnetic field along the c-axis leads to a 
ferromagnetic alignment of the $Gd$-moments along the c-axis. Once the magnetic 
field is switched off the system reverts to its helical AF state. This suggests that
the alloys' $\chi({\bf q})$'s and anisotropies are similar to those shown in Fig.1(b)
and Fig.2(a) but with the relative peak heights of $\chi({\bf q})$ at 
${\bf q}=(0,0,0)$ and ${\bf q}_{inc}$ to be finely balanced.

In {\it conclusion}, we have described an {\it ab initio} theoretical formalism 
to calculate the relativistic static paramagnetic spin susceptibility for metals at 
finite temperatures. Since relativistic effects such as spin-orbit coupling are 
included we can identify the anisotropy or {\it easy axes} of the magnetic response. We applied 
this formalism to the 4d metal $Y$. Its enhanced susceptibility displays a peak at the 
incommensurate wave-vector ${\bf q}_{inc} = (0,0,0.57 \pi/c)$ traceable to a FS 
nesting feature found in the calculated 
electronic structure and fermiology experiments~\cite{Dugdale,Vinokurova}. The 
explanation of magnetic order in rare-earth-$Y$ alloys and superstructures in terms of
the FS topology of $Y$ is well established both on theoretical~\cite{Liu+Yttium_chi}
 and, more recently,
experimental grounds.~\cite{Dugdale, Fretwell} Results in this letter are fully 
consistent with these findings but we have also shown how the electronic structure of
$Y$ influences the canting of the rare earth moments with respect to the crystal axes.
The finding that the incommensurate AF order of these moments has a helical structure
whilst any potential ferromagnetism has an easy-axis parallel to the c-axis is 
suggested as a more general property of the conduction electrons of h.c.p. rare-earth
materials and has a role in the magnetism of RE/$Y$ multilayers. 

We acknowledge support from the EPSRC (UK).
 


\begin{thebibliography}{99}
\bibitem{Remark} We also find this relationship to hold for tetragonal crystal structures.
\end{thebibliography}
\end{document}